\documentclass[prc,twocolumn,showpacs,floatfix,nofootinbib,preprintnumbers,superscriptaddress,amsmath,amssymb]{revtex4}

\usepackage{amsmath}           
\usepackage{amsfonts}
\usepackage{graphicx}          
\usepackage{dcolumn}           
\usepackage{bm}
\usepackage{color}

\begin{document}

\preprint{\fbox{\sc version of \today}}

\title{Surface Symmetry Energy of Nuclear Energy Density Functionals} 
       
\author{N. Nikolov}
\affiliation{
Department of Physics and Astronomy, University of Tennessee, Knoxville, Tennessee 37996, USA
}%
\affiliation{
Physics Division, Oak Ridge National Laboratory, Oak Ridge, Tennessee 37831, USA
}%
\author{N. Schunck}
\affiliation{
Department of Physics and Astronomy, University of Tennessee, Knoxville, Tennessee 37996, USA
}%
\affiliation{
Physics Division, Oak Ridge National Laboratory, Oak Ridge, Tennessee 37831, USA
}%
\affiliation{
Lawrence Livermore National Laboratory,
P.O. Box 808, L-414, Livermore, CA 94551, USA
}%
\author{W. Nazarewicz}
\affiliation{
Department of Physics and Astronomy, University of Tennessee, Knoxville, Tennessee 37996, USA
}%
\affiliation{
Physics Division, Oak Ridge National Laboratory, Oak Ridge, Tennessee 37831, USA
}%
\affiliation{
Institute of Theoretical Physics, University of Warsaw, ul. Ho\.za 69,
PL-00-681 Warsaw, Poland }%
\author{M. Bender}
\affiliation{
Universit\'{e} Bordeaux I, CENBG/IN2P3, Centre d'Etudes Nucl\'eaires de Bordeaux
Gradignan,
UMR5797, Chemin du Solarium, BP120, F-33175, Gradignan, France}
\author{J. Pei}
\affiliation{
Department of Physics and Astronomy, University of Tennessee, Knoxville, Tennessee 37996, USA
}%
\affiliation{
Physics Division, Oak Ridge National Laboratory, Oak Ridge, Tennessee 37831, USA
}%

\date{\today}

\begin{abstract}
We study the bulk deformation properties of  the Skyrme nuclear energy density functionals. Following simple arguments based on the leptodermous expansion and liquid drop model, we apply the nuclear density functional theory to assess the role of the surface symmetry energy in nuclei. To this end, we 
validate the commonly used functional parametrizations against the data on excitation energies of  superdeformed band-heads in Hg and Pb isotopes, and fission isomers in actinide nuclei.  After  subtracting  shell  effects, the results of our self-consistent calculations are consistent with  macroscopic arguments and indicate that experimental data on strongly deformed configurations in neutron-rich nuclei are essential for optimizing future nuclear energy density functionals. The resulting survey provides a useful benchmark for further theoretical improvements. Unlike in nuclei close to the stability valley, whose macroscopic deformability hangs on the balance of surface and Coulomb terms,  the deformability  of neutron-rich nuclei strongly depends on the surface-symmetry energy; hence, its proper determination is crucial for  the stability of deformed phases of the neutron-rich matter and  description of fission rates for r-process nucleosynthesis.
\end{abstract} 


\maketitle


\section{Introduction}
\label{introduction}

Ongoing efforts to develop the nuclear Energy Density
Functional (EDF) of spectroscopic quality  are faced with 
the challenge to find optimal experimental and theoretical constraints that would help us pinning down its various coupling constants. Traditional parameterizations of nuclear EDFs such as Skyrme 
and  Gogny  often rely, through the fitting protocol applied,  on a combination of carefully selected experimental data as well as pseudo data characterizing properties of nuclear matter 
(see, e.g.,  discussion in Refs. \cite{[Ben03],[Sto07b],[Cha08],[Klu09],[Gor10]}). Considering the fairly simple forms (and small number of coupling constants) of commonly used nuclear EDFs, the success of the nuclear  density 
functional theory (DFT) to describe a wide range of nuclear properties has been truly  remarkable. However, the robustness of these 
parameterizations when going away from the neighborhood of the  stability valley, where experimental data are abundant, to the neutron-rich region where data are scarce or non-existent, is questionable.  Indeed,  large 
differences in predictions   for very neutron-rich 
or super-heavy elements seen  for various EDFs \cite{[Dob98c],[Dob02aw],[Erl10]} is highly unsatisfactory. In fact, recent systematic   studies of Skyrme EDFs showed that some coupling constants cannot be properly constrained by existing data, and that the current forms of EDFs are too limiting  \cite{[Ber05],[Kor08]}. Moreover,
early attempts to employ statistical methods of
linear-regression and error analysis \cite{[Fri86]} have been revived
recently and applied to determine the uncertainties
of EDF parameters, errors of calculated observables, and to assess the quality of theoretical extrapolations \cite{[Klu09],[Kor08],[Toi08],[Rei10],[Kor10a]}. The major uncertainty in nuclear
EDF lies in the isovector channels that are poorly constrained by experiment; hence, new data on unstable  nuclei with large neutron excess
having a large ``lever arm" from the
valley of stability  are essential \cite{[RISAC]}.

The pool of fit-observables
entering the optimization protocol of EDF, usually contains
experimental data characterizing both bulk (global)
and local nuclear properties, as well as theoretical pseudo-data pertaining to global  nuclear matter properties (NMP). The characterization in terms of ``bulk" and ``local" is not very precise and somehow arbitrary; it has its origin in the macroscopic-microscopic approach, which offers a description in terms of a macroscopic liquid drop (whose properties change smoothly as a function of nucleon numbers) and shell correction that oscillates rapidly with shell filling \cite{[Bra72],[Bol72],[Boh75w],[Rag95]}. In the context of DFT,  the binding energy of 
a nucleus of mass $A$ and neutron-excess $I = (N-Z)/A$ can  be split 
into a smooth function of $I$ and $A$, and a fluctuating shell correction  term 
by means of  the Strutinsky energy theorem \cite{[Bra72],[Str67a],[Str68],[Rin80],[Bra81]}. This theorem, together with the shell-correction method,  offers a formal 
framework to link the  self-consistent DFT with macroscopic-microscopic models
which often provide useful insights in terms of 
the liquid drop (or droplet) model and shell effects.

Single-particle shell structure can  be accessed through, e.g., 
experimental separation energies and  single-particle strength. Such fit-observables are often used 
in the determination of EDF parameterizations \cite{[Ben03],[Kor10b]}. 
Similarly, the volume and symmetry terms of the liquid drop model (LDM) \cite{[Mye69],[Mye74]} are related 
to nuclear matter properties, which are relatively well determined. This 
effectively constrains specific combinations of parameters of the EDF.
However, surface terms of the LDM  are  harder to pin down. 
In a recent work \cite{[Rei06]}, the leptodermous expansion of the nuclear binding energy was revisited, and the LDM
parameters of several EDFs were extracted from  DFT calculations for very large nuclei. 
While the surface and curvature terms 
came out to be fairly robust, it was found that the surface-symmetry 
coefficient $a_{\text{ssym}}$  of the LDM could vary by a factor 3 across the set 
of parameterizations. This coefficient enters the expression for the symmetry energy in the LDM:  
\begin{equation}\label{Esym}
\frac{E^\text{LDM}_{\text{sym}}}{A}=\left(a_{\text{sym}} + \frac{a_{\text{ssym}}}{A^{1/3}}\right)I^2,
\end{equation}
where $a_{\text{sym}}$ is the (volume) symmetry energy coefficient.
A thorough   compilation of symmetry energy coefficients obtained in different EDFs, assuming various definitions, can be found in Sec. 4.7 of Ref.~\cite{[Dan09]}.

In early Hartree-Fock (HF) and extended-Thomas-Fermi  studies 
\cite{[Far78],[Far80],[Far81],[Ton84]} using various EDFs, a  correlation
between $a_{\text{ssym}}$ and $a_{\text{sym}}$ was pointed out. Namely,  EDFs having large values of $a_{\text{sym}}$ have also large 
$|a_{\text{ssym}}|$. Since the surface-symmetry coefficient is negative, these two terms act in opposite directions  in $E_{\text{sym}}$. The correlation between bulk and surface symmetry energy was further discussed in 
Refs.~\cite{[Dan03],[Ste05],[Die07],[Kir08],[Dan09]}; it was concluded that 
the presence of the correlation makes an absolute   determination
of $a_{\text{sym}}$ and
$a_{\text{ssym}}$ from nuclear masses difficult (see also discussion in an early Ref.~\cite{[Koh76]}).

The experimental information about $a_{\text{ssym}}$ is fairly limited.
The ratio of the surface-symmetry to symmetry (or volume-symmetry) coefficients, $r_{S/V}=a_{\text{ssym}}/a_{\text{sym}}$, has been estimated from 
the electric dipole strength distribution  \cite{[Lip82]},  masses
\cite{[Mol95],[Muk07],[Kir08],[Wan10]},  masses and radii \cite{[Dan03],[Die07]}, and excitation energies of isobaric
analog states \cite{[Dan07]}.
Recently, an attempt has been made to extract $a_{\text{ssym}}$ \cite{[Kol10]} from the separation energies through the displacement of neutron and proton chemical potentials. They noted a large $A$-variation of $r_{S/V}$. 
As discussed later, 
the  DFT values of $r_{S/V}$ obtained in Refs.~\cite{[Rei06],[Sat06],[Dan09]}  are fairly consistent with  phenomenological estimates.

Since the absolute value of $a_\text{ssym}$ is not
well constrained by experimental data on  ground-state (g.s.) nuclear properties, one needs to find some  mechanism  that would enhance the  surface-symmetry term with respect to the  dominant volume symmetry energy. Since the surface-symmetry energy increases with both neutron excess and the nuclear surface area,
it is anticipated that strongly deformed configurations of nuclei with appreciable neutron excess can be of help. Indeed, the nuclear shape deformation increases the surface area thus amplifying the surface-symmetry energy in a neutron-rich nucleus. Conversely, 
the precise determination of surface-symmetry energy is important
to describe the deformability of neutron-rich systems and validate theoretical extrapolations. In this context, one can mention several phenomena involving neutron excess and deformation:
\begin{itemize}
\item
{\it Position of the neutron drip line}. Deformed nuclei are expected in several regions near the neutron drip line \cite{[Sto03],[Gor09]}. In some cases, deformation energy can impact their mere existence. For instance, it has been predicted that there exist  particle-bound even-even nuclei that have at the same time negative two-neutron separation energies due to shape coexistence effects \cite{[Sto03]}.
\item
{\it Borders of the superheavy region}. 
The super- and hyperheavy nuclei with $Z$ $>$126 can exist in states associated with very exotic topologies of nuclear density as the competition
between Coulomb, surface, symmetry, and shell effects can give rise to formation
of voids  \cite{[Won73]}. The subject of
exotic (bubble, toroidal, band-like) configurations in nuclei with very
large atomic numbers has been addressed by several studies
\cite{[Dec03],[Yu00a],[Naz02],[Vin08]}. It is difficult to say at
present whether these exotic  topologies can occur as metastable states
\cite{[Naz02]} and what is their stability to various shape deformations.
\item
{\it Fusion and fission of neutron-rich nuclei.} 
Synthesis of heavy and superheavy neutron-rich nuclei is profoundly affected by nuclear deformability through the energetics of fusion and fission valleys \cite{[Mol04],[Zag08]}. Our ability to describe fission of neutron-rich systems is also important for modeling nuclear reactors.
While DFT calculations are currently able to predict barrier heights of known nuclei with a typical accuracy of  20\%,  the resulting uncertainties in  fission cross sections  are still large \cite{[Gor09]}. The ability of modern nuclear EDFs to predict neutron induced-fission rates for neutron rich nuclei that cannot be measured is crucial. At this point, the dependence of rates on fission barriers is appreciable \cite{[Pan10]}.
\item
{\it Rotational properties of neutron rich nuclei.}
Nuclear deformation determines the  response of the nucleus to angular momentum.
Little is known about the collective rotation of very neutron-rich systems \cite{[Naz01a]} and the corresponding interplay between deformation, isospin, and rotation.
\item
{\it Astrophysical r-process.} Fission of neutron rich nuclei impacts the formation of heavy elements at the final stages of the r-process through the recycling mechanism \cite{[Cam02],[Pan08]}. The fission recycling is believed to be of particular importance during   neutron star mergers where free neutrons of high density are available \cite{[Pan08],[Pan10]}. Also, neutrino-induced fission of neutron rich nuclei could affect the r-process flow in some scenarios \cite{[Kol04]}.
\item
{\it Structure of neutron stars.} Nucleonic phases in the inner crust of neutron stars are associated with very neutron-rich deformed nuclei or strongly deformed pasta and anti-pasta phases \cite{[Mag02],[New09]}.
\end{itemize}

The primary motivation  of this paper is to
assess the role played by  the surface-symmetry energy in neutron rich nuclei.  
To this end, we study the  surface-symmetry contribution to the LDM energy and demonstrate that it may be as important as the Coulomb term in  driving deformation 
properties of very neutron-rich systems. Using the self-consistent DFT, we survey 
excitation energies of  superdeformed (SD) states in the Hg-Pb region and SD fission isomers in the actinides.  By  subtracting  microscopic shell  corrections, we extract the macroscopic part of the HF deformation energy and demonstrate that the outcome is consistent with  qualitative macroscopic estimates. Our results indicate that experimental data on strongly deformed configurations in neutron-rich nuclei are key for optimizing the isospin channel of the nuclear energy density functional.

Our paper is organized as follows. In Sec. \ref{Sec:assym}, based on the spherical and deformed LDM, we discuss general properties of the surface-symmetry term. In particular,  we analyze    symmetry-energy parameters  of various  Skyrme EDFs,  study global correlations between $a_{\text{ssym}}$ and $a_{\text{sym}}$, and show how to disentangle the surface-symmetry term by studying deformed configurations in nuclei with nonzero neutron excess.
Self-consistent  DFT calculations 
of fission isomers in the actinides and band-heads
of super-deformed rotational bands in the  $A\sim 190$ region are 
presented in Sec. \ref{Sec:HFB}. The methodology used to extract the smooth 
contribution to the total energy is outlined in  Sec. \ref{Sec:leptodermous}.
Section~\ref{results}   presents our calculations of the smooth contributions to the deformation energy of SD states and compares HFB and LDM results. 
Finally, the conclusions of our work are given in Sec.~\ref{conclusion}.


\section{Liquid Drop Model from the  Skyrme Energy Density Functional}
\label{Sec:assym}

The LDM provides a convenient parametrization of the bulk part of the binding energy of a spherical even-even
nucleus with $Z$ protons and $N$ neutrons. Expressed in terms of volume ($a_{\text{vol}}$), surface ($a_{\text{surf}}$), curvature ($a_{\text{curv}}$), symmetry, surface-symmetry, Coulomb, and Coulomb exchange parameters, it reads:
\begin{multline}
E_{\text{sph}}^{\text{LDM}} = a_{\text{vol}}A + a_{\text{surf}}A^{2/3} + a_{\text{curv}}A^{1/3} \\
+ a_{\text{sym}}I^{2}A + a_{\text{ssym}}I^{2}A^{2/3} + a^{(2)}_{\text{sym}}I^{4}A \\
+ \frac{3}{5}\frac{e^{2}}{r_{0}^{\text{ch}}}\frac{Z^{2}}{A^{1/3}} 
- \frac{5}{4} \left( \frac{3}{2\pi}\right)^{2/3}
\frac{3}{5}\frac{e^{2}}{r_{0}^{\text{ch}}}\frac{Z^{4/3}}{A^{1/3}},
\label{Eq_LDM_sph}
\end{multline}
where   $e$ is the electric charge and $r_{0}^{\text{ch}}$ the Wigner-Seitz radius. The justification of (\ref{Eq_LDM_sph}) can be given in terms of the  leptodermous expansion valid for systems with a well developed surface \cite{[Vin75],[Blo77],[Gra82]} that sorts the various contributions to the binding energy of finite nuclei in terms that have transparent physical meaning.
The expansion (\ref{Eq_LDM_sph})  can be extended to higher orders \cite{[Mye98b],[Liu02a]}. The second-order symmetry energy term $\propto I^{4}$ 
is not always included in the macroscopic LDM but it  naturally present in the microscopic LDM expression \cite{[Rei06]}.


\subsection{Approaches to Bulk Nuclear Properties}
\label{Subsec:bulk}

Some LDM parameters are fundamental NMPs and can be determined microscopically from  ab-initio calculations of the equation of state of nucleonic matter \cite{[Akm98],[Sam10],[Dal10]}. Another, phenomenological strategy is to obtain LDM constants, or at least some of them, from
a direct fit to selected experimental data from 
finite nuclei. The original work of Myers and Swiatecki followed such a 
strategy \cite{[Mye66]}: by modeling  local  fluctuations in particle numbers due to shell effects, one can extract smooth LDM trends 
from experimental nuclear masses. Subsequent refinements involved the upgrade 
from a simple drop to a more accurate droplet model \cite{[Mye74]}, which 
allowed to pin down additional terms in the leptodermous expansion. 
Further refinements can be found in, e.g.,  Refs. \cite{[Mye82],[Mol95],[Mye98b],[Mye98c],[Pom02]}.

Just as in the microscopic  approach, whose outcome depends on both the input (i.e., nucleon-nucleon interactions) and the theoretical method used to solve the many-body problem,
the results of phenomenological procedure depend on the choice of fit-observables and prescription used  to compute  shell corrections (see  Ref.~\cite{[Sal10]} for  a recent concise overview of this topic). 
There are significant correlations among the different LDM terms and some parameters are poorly determined \cite{[Kir08]}. In particular, precise extraction of higher-order  isospin-dependent terms 
requires  abundant data for very neutron-rich and/or heavy nuclei, which are not  available at present.


\subsection{Spherical Liquid Drop Based on Density Functional Theory}
\label{Subsec:sphLDM}

An  advantage of the macroscopic approach to bulk nuclear properties is that 
it can also be applied in the context of the nuclear DFT.
While some LDM constants pertaining to infinite or semi-infinite NMP can  be extracted from EDF parameterizations \cite{[Ben03],[Rik03],[Dan09]}, surface and curvature terms
are best determined by  using the semi-classical approach \cite{[Far85],[Cen98a]} or by removing the contribution from shell effects 
from self-consistent DFT  results \cite{[Bra81]}. 
There are relatively few examples of latter studies in the literature, and most 
were confined to spherical symmetry. In Ref.~\cite{[Kle02]}, the parameters 
$a_{\text{vol}}$, $a_{\text{sym}}$, $a_{\text{surf}}$, $a_{\text{ssym}}$ and 
$r_{0}^{\text{ch}}$ were estimated from spherical HFB calculations using the 
finite-range Gogny force D1S \cite{[Ber91b]}. In 
\cite{[Pom02a]}, a similar work was carried out for the NL3 parametrization 
of the Relativistic Mean Field (RMF) \cite{[Lal97]}.

This program was  carried out more systematically in Ref.~\cite{[Rei06]} for  Skyrme EDFs and several parameterizations of RMF Lagrangians using the HF method. 
The main difference with \cite{[Kle02],[Pom02a]} is that the convergence of the 
leptodermous expansion was tested using an extended sample of very large spherical nuclei. The Coulomb terms
were ignored to be able to approach nuclei of arbitrary sizes and to avoid radial instabilities (Coulomb frustration) characteristic of systems with many protons. The shell corrections were extracted from the 
HF results according to Green's function method and the generalized Strutinsky smoothing procedure of Refs.~\cite{[Kru00b],[Ver00],[Cwi05a]}. 
Table~\ref{table01} displays the values of the symmetry, surface-symmetry 
and surface coefficients for various realizations of the nuclear EDF. These values are fairly close to those obtained in Ref.~\cite{[Dan09]}. We also note
that in the case of  NL3, there are relatively large differences 
between the values in Table \ref{table01} and those reported in Ref.~\cite{[Pom02a]}: 
these can be attributed to different ways of extracting the shell correction, 
and fitting the LD formula (sample size, treatment of the 
Coulomb term). 

\begin{table}[ht]
\begin{center}
\caption{
Surface, symmetry,  and surface-symmetry LDM coefficients (in MeV)  of various EDFs extracted from leptodermous expansion in Ref.~\cite{[Rei06]}. The Skyrme EDF parametrizations are:
SkM* \cite{[Bar82]}, SkP \cite{[Dob84]}, BSk1 \cite{[Sam02]}, BSk6 
\cite{[Gor03]}, SLy4-SLy6 \cite{[Cha98]}, SkI3-SkI4 \cite{[Rei95]}, SkO 
\cite{[Rei99]}. The RMF Lagrangians are: NL1 \cite{[Rei86]}, 
NL-Z \cite{[Ruf88]} and NL-Z2 \cite{[Ben99a]}. For comparison, the results of LDM fits are given:
LDM$^{(1)}$ \cite{[Mol95]} and  LDM$^{(2)}$-LSD \cite{[Pom02]}.
}
\begin{ruledtabular}
\begin{tabular}{cccc|cccc}
EDF & $a_{\text{sur}}$ & $a_{\text{sym}}$ & $a_{\text{ssym}}$ &  EDF & $a_{\text{sur}}$ & $a_{\text{sym}}$ & $a_{\text{ssym}}$  \\
\hline
SkM* & 17.6  & 30.04 & -52 &  NL1  & 18.8  & 43.48 & -110  \\
SkP  & 18.2  & 30.01 & -45 &  NL3  & 18.6 & 37.40 & -86    \\
BSk1 & 17.5 & 27.81 & -36  &  NL-Z & 17.8 & 41.74 & -125   \\
BSk6 & 17.3 & 28.00 & -33  &  NL-Z2& 17.4 & 39.03 & -90   \\
SLy4 & 18.4 & 32.01 & -54  &       &       &       &  \\
SLy6 & 17.7 & 31.96 & -51  &  LDM$^{(1)}$  & 21.1 & 30.56 & -48.6  \\
SkI3 & 18.0 & 34.84 & -75  &  LDM$^{(2)}$  & 19.4 & 29.28 & -38.4  \\
SkI4 & 17.7 & 29.51 & -34  &  LSD  & 17.0  & 28.82 & -38.9  \\
SkO  & 17.3 & 31.98 & -58  &       &       &       & 
\end{tabular}
\label{table01}
\end{ruledtabular}
\end{center} 
\end{table}

As noted in Ref.~\cite{[Rei06]}, the leading surface and symmetry terms appear 
relatively similar within each family of EDFs, with a clear difference for $a_{\text{sym}}$ between  non-relativistic and relativistic approaches. Obviously,  even  small variations in $a_{\text{surf}}$ and $a_{\text{sym}}$ seen in Table \ref{table01} can have an appreciable impact on the binding energy,  as these coefficients are multiplied by large $A$- and  $I^2$-dependent factors \cite{[Rei06]}. For the surface-symmetry coefficient, however, there are  much larger discrepancies. For Skyrme EDFs, for instance, there is a factor of two between the largest and smallest value. This  demonstrates that $a_{\text{ssym}}$ is very poorly constrained in the current EDF parameterizations (see also discussion in \cite{[Dan09]}).
\begin{figure}[htb]
\includegraphics[width=\linewidth]{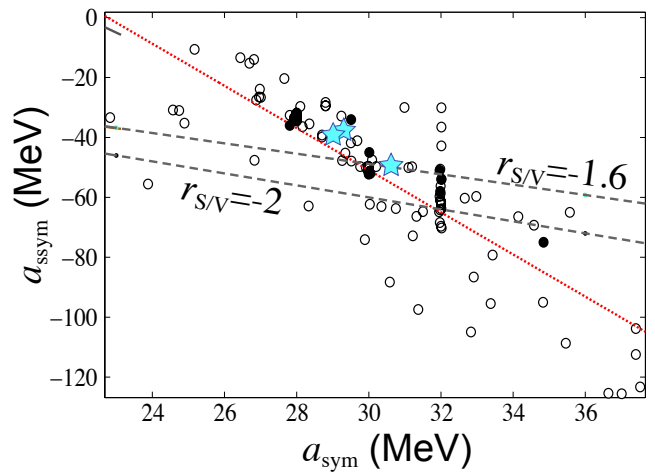}
\caption{
	     (color-online) Correlation  between the symmetry and surface-symmetry 
	     coefficients  extracted from  Skyrme EDFs from Table~\ref{table01} (dots) and Skyrme EDFs of Ref.~\cite{[Dan09]} (circles).  The phenomenological LDM values of Table~\ref{table01} are also indicated (stars) as well as  the  hydrodynamical \cite{[Lip82]} and mass  \cite{[Sat06],[Kol10]} estimates:   $r_{S/V}= -2$ and $-1.6$, respectively (dashed lines). The linear fit to  the values of Ref.~\cite{[Dan09]} is shown by a dotted line.
	    }	 
\label{fig01}	 
\end{figure}

In addition, as discussed in the previous section, there appears a 
clear (anti-)correlation between the (positive) value of the symmetry coefficient and the 
(negative) value of the surface-symmetry coefficient \cite{[Far78],[Far80],[Far81],[Ton84],[Dan03],[Ste05],[Die07],[Kir08],[Dan09]}. Figure~\ref{fig01}, 
displays the pairs $(a_{\text{ssym}}, a_{\text{sym}})$ for the  Skyrme EDFs 
of Table \ref{table01} and  the EDFs of Ref.~\cite{[Dan09]}. 
The ratio $r_{S/V}$ extracted from experimental masses is $r_{S/V}\approx -1.7$ \cite{[Kir08]}. When looking into details, however, it  exhibits a large $A$-variation \cite{[Kol10]}; $r_{S/V}$
ranges between $-1$ (for $A\ge 12$) and $-1.7$ (for $A\ge 50$).
As discussed in \cite{[Sat06]}, the data on the electric dipole polarizability 
yields $r_{S/V}\approx -1.65$. 
The  DFT values shown in  Fig.~\ref{fig01} are
not inconsistent with these phenomenological estimates. While a correlation between  $a_{\text{sym}}$ and
$a_{\text{ssym}}$ is clear,
a very large spread of values is indicative of the inability of current data on 
g.s.  nuclear properties to adequately constrain $a_{\text{ssym}}$. It is interesting to note that the LDM values of Table~\ref{table01} and phenomenological estimates cluster around $a_{\text{sym}}=30$\,MeV and
$a_{\text{ssym}}=-45$\,MeV.

\begin{figure}[!]
\includegraphics[width=\linewidth]{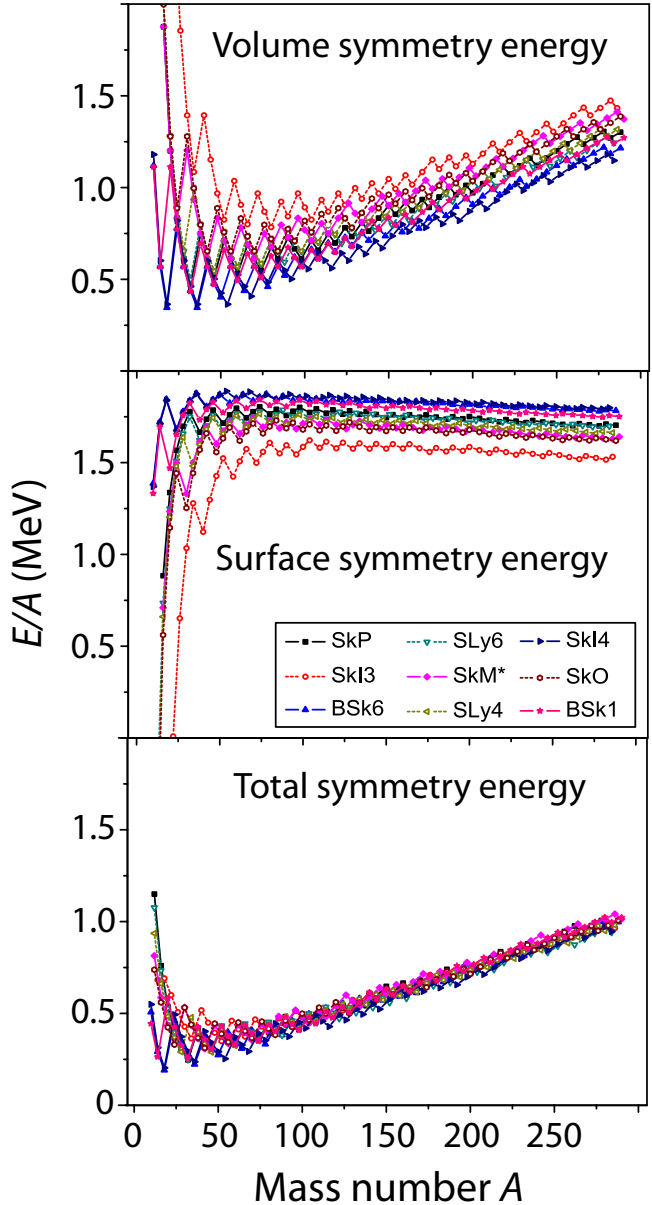}
\caption{
     	(color-online)  Contributions to the microscopic LDM
     	energy per nucleon along the LDM valley of stability: volume symmetry term 
		$a_{\text{sym}}I^{2}$ (top), 
		surface-symmetry term $a_{\text{ssym}}I^{2}A^{-1/3}$ (middle), and the total symmetry energy (bottom),  for the microscopic LDM derived from  Skyrme EDFs  of Table~\ref{table01}.
	    }	 
\label{fig02}
\end{figure}

To get more insights into the consequences of this observation, we
plot in Fig. \ref{fig02} the symmetry and surface-symmetry contributions to the binding energy per nucleon $E^{\text{LDM}}/A$ for the microscopic LDM derived from  Skyrme EDFs 
listed in Table~\ref{table01}  along the LDM valley of stability. The latter one is defined by minimizing the LDM energy
in the  $(Z,N)$ plane. The symmetry energy slightly increases with mass number 
(upper panel), owing to the fact that the valley of stability bends 
down for increasing $Z$, so that the energy per nucleon $E_{sym} = 
a_{\text{sym}}I^{2}$ increases. Contrariwise, the surface-symmetry 
energy slightly decreases with $A$  because of the $A^{-1/3}$ dependence.

Interestingly, the total symmetry energy (i.e., the sum of volume and surface  contributions) exhibits 
much smaller spread between various EDFs: from about 
$0.4/A$ MeV for both $E_{\text{sym}}/A$ and $E_{\text{ssym}}/A$ to about 
$0.1/A$ MeV for the sum. This is a consequence of the aforementioned correlation between volume and surface symmetry energies that implies that large discrepancies between  individual contributions to the bulk energy tend to cancel out at 
the level of the total binding energy. We note that nuclear binding energies are indeed prime fit-observables constraining parameters of most EDFs.


\subsection{Deformation Energy of Nuclear Liquid Drop}
\label{Subsec:defLDM}

The relative LDM contributions of the volume-symmetry and surface-symmetry energies can be  disentangled if shape deformation is present.  Indeed, the 
 deformation  energy of the deformed LDM can be written as \cite{[Bol72],[Has88],[Mol95]}:
\begin{multline}
E^{\text{LDM}}_{\text{def}} = E^{\text{LDM}} - E_{\text{sph}}^{\text{LDM}}=\\
 (b_{\text{s}}-1)a_{\text{surf}}A^{2/3} + (b_{\text{curv}}-1)a_{\text{curv}}A^{1/3} + \\
(b_{\text{s}}-1)a_{\text{ssym}}I^{2}A^{2/3} +
(b_{\text{c}}-1)\frac{3}{5}\frac{e^{2}}{r_{0}}\frac{Z^{2}}{A^{1/3}},
\label{Eq_LDM_def}
\end{multline}
where the  geometrical factors $b_{\text{s}}$, $b_{\text{curv}}$ and $b_{\text{c}}$ depend on the shape  of the deformed drop (by definition, they are equal to unity at the spherical shape).
Since the nuclear volume is conserved in the LDM, the surface and curvature $b$-factors increase with deformation. On the other hand, $b_{\text{c}}$ becomes less than one as the Coulomb energy of the deformed drop is lower than that of the spherical drop. While the 
volume-symmetry energy is shape-independent, the surface-symmetry term has the same dependence on the nuclear shape and $A$ as the surface term. Consequently these two contributions to the symmetry energy behave  differently in deformed nuclear drops.  
\begin{figure}[htb]
\centering
\includegraphics[width=\linewidth]{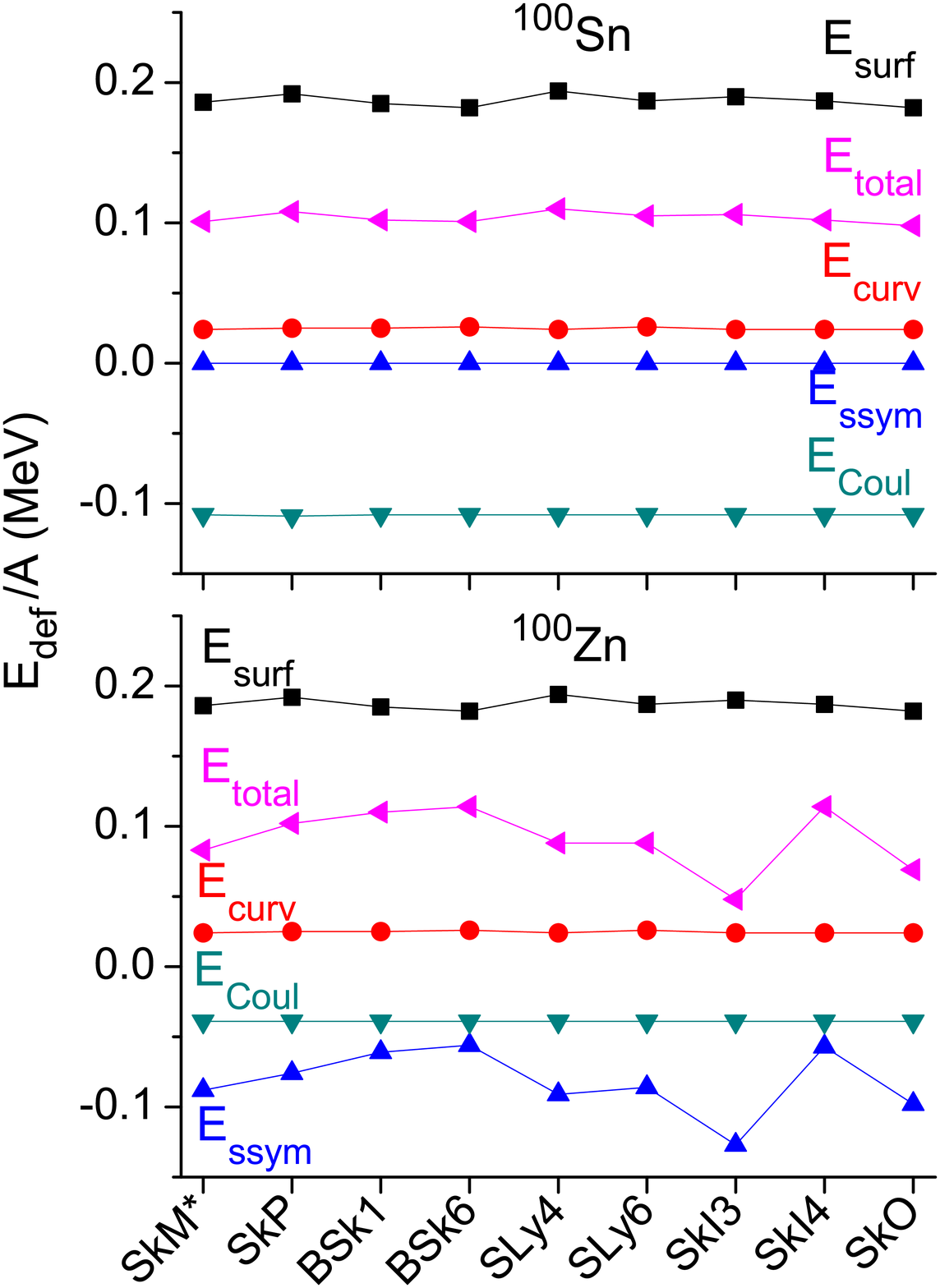}
\caption{
		 (color online) Individual contributions to the total deformation 
		 energy per nucleon of  the microscopic LDM  for nine Skyrme EDFs for two  $A=100$ isobars:  $^{100}$Sn ($I=0$, top) and  $^{100}$Zn ($I=0.4$, bottom). The assumed  quadrupole  deformation is $\tilde\beta_2 = 0.6$.}	 
\label{fig03}	 
\end{figure}

As an example, we plot in Fig.~\ref{fig03} the 
individual contributions to 
$E^{\text{LDM}}_{\text{def}}$
for the two  $A=100$ drops at a fixed quadrupole deformation $\tilde\beta_2 = 0.6$. (All  remaining  deformations are set to zero.)  Specifically, shown are contributions from  the surface, 
curvature, Coulomb, and surface-symmetry terms corresponding to different Skyrme EDFs. The Coulomb radius $r_0$  was assumed to be the same as
the Wigner-Seitz radius defining the saturation density.

Since for  $^{100}$Sn the isospin excess is zero,  the deformation energy contribution coming 
from the surface-symmetry term vanishes. The 
variations  of the total LDM energy between different EDF parametrizations
are relatively small and  primarily related to slightly different values of $a_{\text{surf}}$.
The picture changes dramatically  when going to $^{100}$Zn, a very 
neutron rich nucleus with isospin $I = 0.4$. The  variations  between predictions of different EDFs have a much 
larger amplitude and are caused almost exclusively by the surface-symmetry 
term. This indicates that the theoretical differences in the LDM deformation 
energy of heavy neutron-rich nuclei are almost 
entirely driven by the poorly determined surface-symmetry term. As it is clear from Fig.~\ref{fig03}, an experimental access to this term can be provided by extracting shell energy from the measured masses of very deformed configurations in neutron-rich nuclei. Another interesting observation drawn from the deformed LDM exercise is that, contrary to the usual scenario in which the macroscopic deformability is solely driven by the competition between  surface and Coulomb terms, the macroscopic  deformation energy of very neutron rich nuclei 
involves a three-way competition between the repulsive surface term and attractive Coulomb and surface-symmetry terms.


\section{Skyrme HFB calculations at Large Deformations}
\label{Sec:HFB}

The binding energy of a deformed nuclear configuration can be decomposed into a macroscopic part and shell correction. In order to determine whether the macroscopic features related to the surface-symmetry energy, identified in Sec.~\ref{Subsec:defLDM}, show up in self-consistent DFT calculations for actual nuclei, we have performed Hartree-Fock-Bogoliubov (HFB) calculations for a number of states at large deformation with available experimental information.

\subsection{Survey of Superdeformed Bandheads and Fission Isomers}

We selected two regions of the nuclear chart: in the actinides, there is a number of isotopes where the excitation energy of the fission isomer is relatively well-known \cite{[Sin02]}. In the neutron-deficient Hg and Pb isotopes, the linking transition between the SD and g.s. bands have been identified for several nuclei, so that the energy of the $0^{+}$ band-head could be extracted \cite{[Hen91],[Hen94s],[Lop96],[Hau97],[Wil03],[Wil05]}. All SD band-head data used in this work are listed in Table.~\ref{tableexp}.
%
\begin{table}[htb]
\begin{center}
\caption{Experimental energies of  $0^{+}$ band-heads of SD states in 
$A$=190 mass region and in the actinides.
}
\begin{ruledtabular}
\begin{tabular}{ccc}
Nucleus & E$_\text{SD}$(0$^+$) (MeV)  & Reference \\
\hline 
{$^{192}$Hg} &  5.3 (9)  & \cite{[Lop96]}\\
{$^{194}$Hg }&  6.017  & \cite{[Hen94s]} \\
{$^{192}$Pb }&  4.011  & \cite{[Hen91]} \\
{$^{194}$Pb} &  4.643  & \cite{[Hau97]}\\
{$^{196}$Pb} &  5.630(5)  & \cite{[Wil05]} \\
$^{236}$U & 2.750  & \cite{[Sin02]}  \\
$^{238}$U &  2.557  & \cite{[Sin02]} \\
$^{240}$Pu &  2.800  & \cite{[Sin02]}  \\
$^{242}$Cm &   1.900  & \cite{[Sin02]}
\end{tabular}\label{tableexp}
\end{ruledtabular}
\end{center}
\end{table}

HFB calculations were performed with the DFT solvers  HFBTHO  \cite{[Sto05]} and
HFODD  \cite{[Dob95],[Dob97],[Dob04],[Dob05],[Dob09d]}. To benchmark binding energies
of superdeformed configurations, we  employed several Skyrme EDFs in the particle-hole channel.
Pairing correlations were
modeled by an mixed-pairing interaction with a  dependence on the isoscalar density \cite{[Dob95c],[Dob02c]}.
All calculations were performed with a cut-off energy of $E_{cut} = 60$ MeV
to truncate  the quasi-particle space. For each parametrization of the Skyrme interaction, the pairing strength 
was fitted to the average neutron pairing gap in $^{120}$Sn 
according to the procedure described in Ref.~\cite{[Dob95c]}. 
In both solvers, the quasi-particle solutions to the HFB problem are 
expanded on the deformed Harmonic Oscillator (HO) basis. Since 
we are probing very elongated systems, we performed the calculations using a stretched basis with 
a large number of  deformed HO shells, $N_\text{max} = 20$.
All calculations were performed assuming axial, reflection symmetric shapes.
The constrained minimization  was performed using the augmented Lagrangian method \cite{[Sta10]} and the procedure of Refs.~\cite{[Dec80],[You09]}.

\begin{figure}[!]
\includegraphics[width=\linewidth]{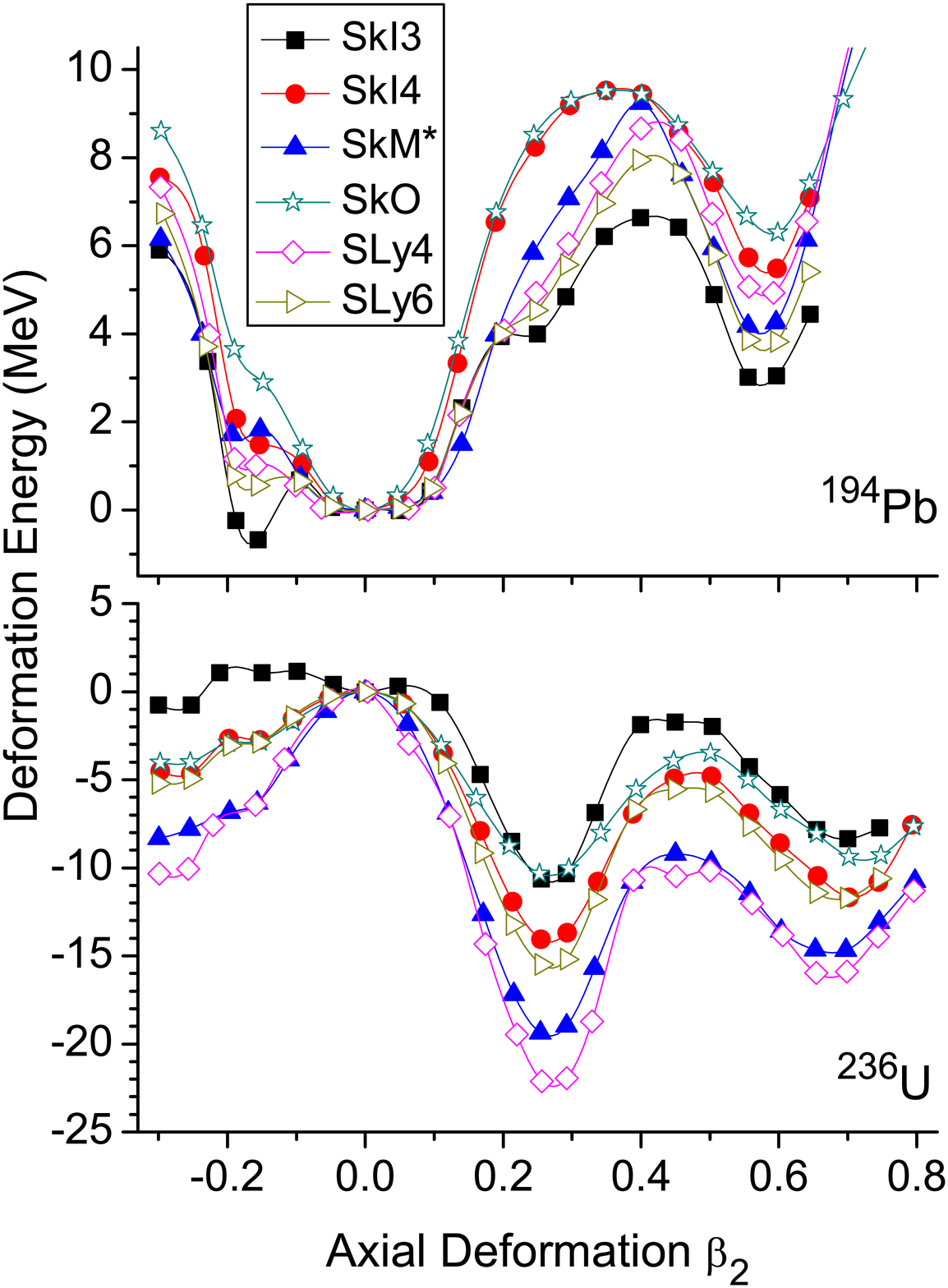}
\caption{
	     (color-online) Potential energy curves for 
	     $^{194}$Pb (top) and  $^{236}$U (bottom) versus quadrupole deformation $\beta$ calculated
	     with  SkI3, SkI4, SkM$^{*}$, SkO, SLy4 and SLy6 Skyrme EDFs.
	     All curves are normalized to the spherical point. Axial symmetry is assumed.
	    }	 
\label{fig04}	 
\end{figure}

As an illustration of typical deformation landscapes in the two regions, Fig. 
\ref{fig04} shows the calculated potential energy curves for $^{194}$Pb and $^{236}$U as functions of 
the  quadrupole deformation $\beta$  extracted from the mass quadrupole 
moment $\langle\hat{Q}_{20}\rangle$ and the total rms radius:
\begin{equation}\label{betadef}
\beta\equiv \sqrt {\frac{\pi}{5}}{\frac{\langle\hat{Q}_{20}\rangle}{\langle r^2\rangle}}.
\end{equation}
While 
the actinide nuclei of interest are always predicted to have    prolate-deformed 
ground states  with  $\beta_{2} \approx 0.3$, 
neutron-deficient Hg and Pb isotopes show a more complex g.s. pattern involving coexisting oblate and spherical structures \cite{[Naz93a]}.

\begin{figure}[h]
\includegraphics[width=\linewidth]{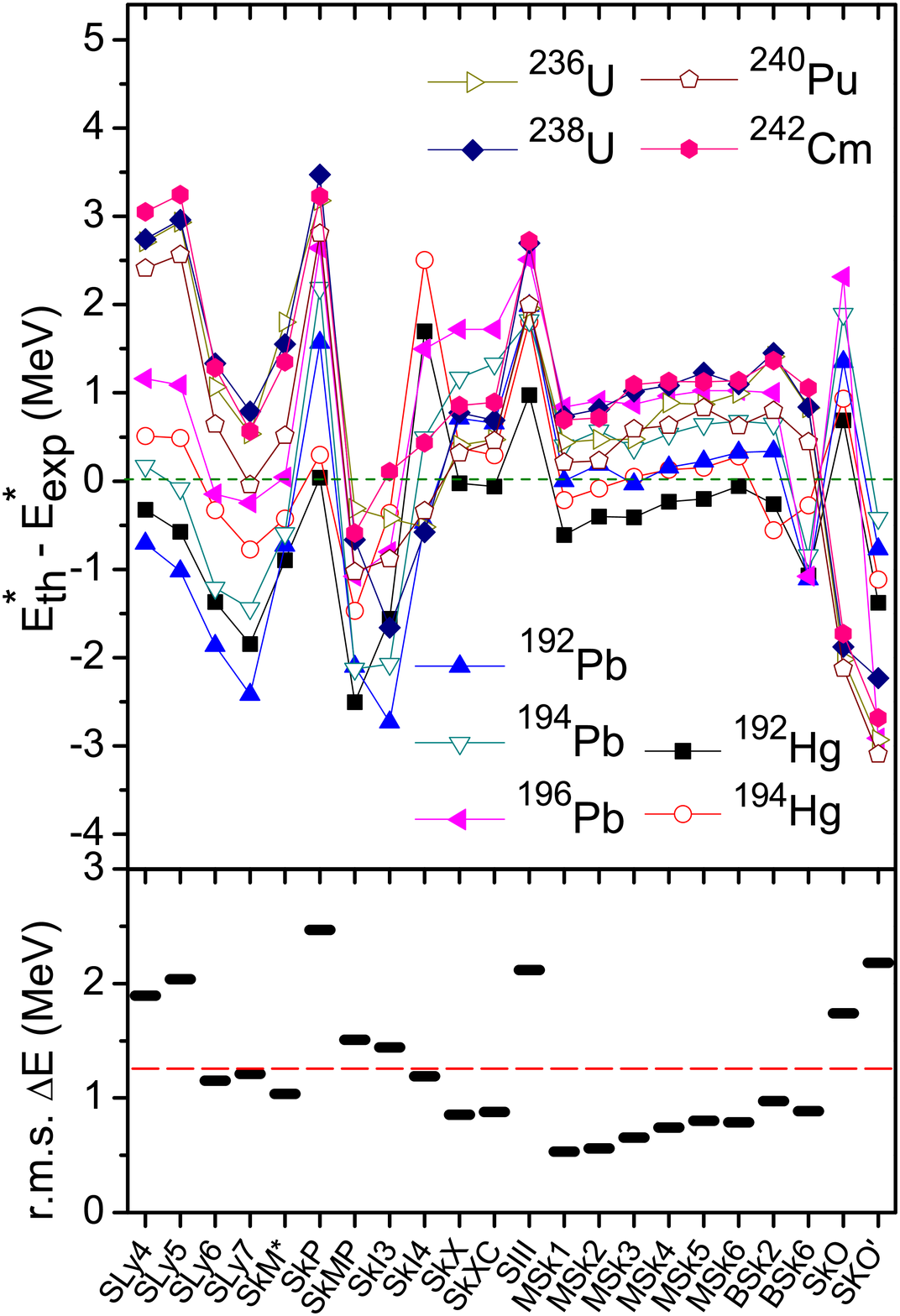}
\caption{
	     (color-online) Residuals $\Delta E = E^{*}_{\text{th}} 
- E^{*}_{\text{exp}}$ (top) and rms deviations from experiment  (bottom) for various Skyrme EDFs.
	     Additional references for the Skyrme forces: SLy5-SLy7 \cite{[Cha98]}, 
	     SkMP \cite{[Ben89d]}, SkX-SkXC \cite{[Bro98a]}, SIII \cite{[Bra95]}, 
	     MSk1-MSk6 \cite{[Ton00]}, BSk2 \cite{[Gor03]}, SkO' \cite{[Rei99]}.
	     The average rms deviation for the set of EDFs considered (marked by a dashed line in the lower panel) is 1.26\,MeV for the nine data points of Table~\ref{tableexp} and 1.34\,MeV for the fission isomers nuclei. 
	    }	 
\label{fig05}	 
\end{figure}

The predicted excitation energy of the SD minimum relative to the g.s.,
$E^{*}_{\text{th}} = E_{\text{SD}} - E_{\text{g.s.}}$,
between the HFB SD minimum  and g.s. minimum 
can be compared with the experimental value $E^{*}_{\text{exp}}$. 
The residuals  $\Delta E = E^{*}_{\text{th}} 
- E^{*}_{\text{exp}}$ are plotted in Fig.~\ref{fig05} for 22 different Skyrme EDFs. It is rather striking 
to notice that, for a given nucleus, the differences between various EDFs   
can be as high as 4 MeV, which is  often greater
than the excitation energy itself. These large fluctuations sometimes 
 occur within a  family of Skyrme EDFs, e.g.,  SLy[x], and has been explained in some cases by the different recipes to treat the center of mass \cite{[Ben00d]}. By contrast, the Brussels-Montreal parametrizations Bsk[x] and Msk[x] are more consistent 
with one another.  An appreciable  EDF-dependence  for SD states had already been pointed out in previous Refs. \cite{[Tak98],[Hee98]}.  Similarly, the sensitivity of fission barriers on EDF parametrizations was studied  in  Refs.~\cite{[Dut80],[Ton85],[Bur04]}. In the context of this work, it is especially interesting to point out that the surface-symmetry term  has been claimed \cite{[Ton85]} to have a significant  influence on self-consistent fission barriers.

The large spread in calculated values of $E^*$ comes as little surprise:  
very few EDFs have been optimized by considering data probing large deformations. 
The importance of considering strongly deformed shapes when fitting the coupling constants of EDFs was discussed in Refs.~\cite{[Gue80],[Bar82],[Ton84]} but this program has been carried out only in a handful of cases.
The SkM* functional \cite{[Bar82]} has been fitted by considering the experimental 
fission barrier in  $^{240}$Pu. The D1S Gogny interaction \cite{[Ber84],[Ber91b]} was also optimized for fission properties.  
In the Bsk14 EDF of the HFB-14 mass model \cite{[Gor07]}, data on fission barriers were utilized to optimize the EDF parameters by adding a phenomenological collective correction accounting for the zero-point rotational-vibrational motion. In this article, we do not employ zero-point corrections as we are primarily interested in the deformation properties of the functionals themselves. 
We refer, e.g., to \cite{[Ben04c]} for a more thorough discussion of dynamical correlations  and their impact on deformation properties of nuclei. We note in passing that such correlations are supposed to impact standard DFT predictions of g.s. energies of Hg and Pb nuclei due to coexistence effects \cite{[Bon90c],[Ben04b]}. 

\subsection{Estimation of Theoretical Errors}

\begin{figure}[h]
\includegraphics[width=\linewidth]{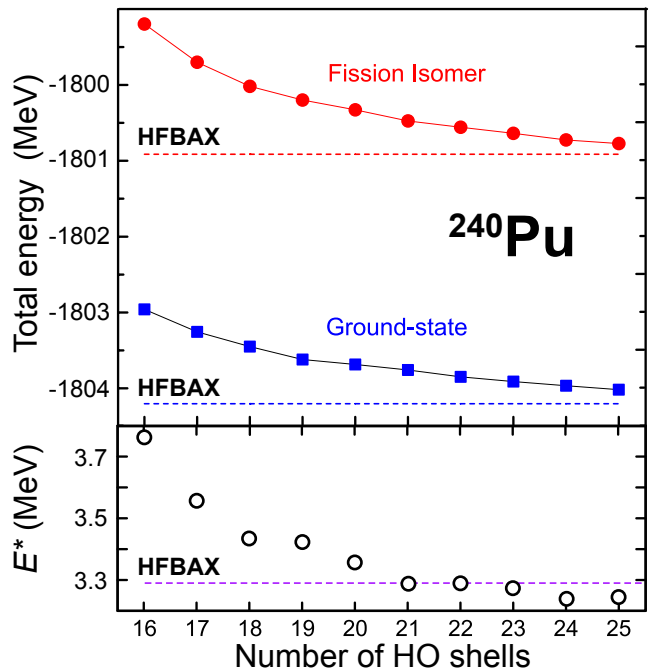}
\caption{
	     (color-online) Convergence of the HO basis expansion for the HFB+SkM* binding energy of  fission isomer  and ground state (top panel) and  excitation energy of fission isomer (bottom panel) in $^{240}$Pu 
		 as a function of the HO basis size. Results are compared with the benchmark numbers  obtained with the precise 
		 coordinate-space solver HFBAX \cite{[Pei08a]}. 
	    }	 
\label{fig06}	 
\end{figure}

Since the values of $\Delta E$ in Fig.~\ref{fig05} are subject to
numerical and experimental uncertainties, it is important to estimate
their respective errors before assessing the model dependence of
results. In order to validate $E^{*}_{\text{th}}$, we studied the
convergence of our HFB results with respect to the size of HO space
used. Figure~\ref{fig06} shows the HFB+SkM* energy of the g.s. and SD
state of $^{240}$Pu calculated with the  HFBTHO solver as a function of
$N_\text{max}$, The HFBTHO numbers are compared to the benchmark results
obtained with the precise coordinate-space DFT solver  HFBAX
\cite{[Pei08a]}.

With the large HO basis used here ($N_\text{max}$=20), the theoretical 
error on the energy of either the g.s. or the SD state is around
600 keV.  Since the HFB theory is variational, the error on the excitation energy is 
in fact much smaller (see the bottom panel of Fig.~\ref{fig05}), and comes principally from the different convergence 
rates of HFB states with $\beta_{2} \approx 0.3 $ and $\beta_{2} \approx 0.6 $. 
Those differences are due to the combination of effects coming from the basis 
deformation  and the choice of oscillator frequency $\hbar\omega$. At $N_\text{max} \geq 16$, 
the dependence on  $\hbar\omega$ is rather weak; hence, the only remaining source of fluctuations  is  the basis deformation. For the residuals, we estimate the latter empirically to be at most 500 keV for 16 shells, and less than 100 keV for 
$N_\text{max} \geq 16$ shells.

The convergence pattern of HFB calculations seen  in Fig. \ref{fig06} is to a large extent exponential. A similar behavior has  also been observed in ab-initio calculations of Refs.~\cite{[Hor99],[Mar09a],[Hag07]}. However, in all these many-body approaches, the size of the actual model space grows combinatorially with the number of active particles and single-particle states taken,  which is not the case for DFT. In a recent work, the exponential convergence of  wave-functions expanded  in a HO basis has in fact been related to its weak differentiability properties \cite{[Kva09]}. It has been  argued  therein that this may be a generic property of systems with exponentially decaying wave-functions. 

For all the   nuclei considered in Fig. \ref{fig06} and Table~\ref{tableexp}, experimental g.s.
masses are known to an excellent  precision of approximately 2 keV
\cite{[Aud03]}. In the  $A\sim 190$ region, the uncertainty of the SD
band-head comes from the extrapolation of the  rotational band down to 
spin $0^{+}$. This procedure is slightly model-dependent, but its error
is estimated to be  only $\sim$5 keV
\cite{[Hen91],[Hen94s],[Lop96],[Hau97],[Wil03],[Wil05]}. In the
actinides, the determination of the excitation energy of the fission
isomer is slightly less  precise:  it is about 5-10 keV for
$^{236,238}$U and about 200 keV for $^{240}$Pu and $^{242}$Cm
\cite{[Sin02]}.

Considering the above, the  theoretical fluctuations of $\sim$several  MeV in $\Delta E$ seen in Fig. \ref{fig05} are well above  numerical uncertainties in $E^{*}_{\text{th}}$ and experimental uncertainties  in $E^{*}_{\text{exp}}$. Consequently, these deviations are rooted in actual EDF parametrizations. In the next section, we shall investigate the relation between the fluctuations in $\Delta E$ and the underlying LDM description.


\section{Bulk Deformation Energy of the Skyrme Energy Density Functional}
\label{Sec:leptodermous}

In order to extract the smooth LDM energy from HFB results, the fluctuating contributions to the energy (i.e., shell-correction  and pairing terms) must be removed. After describing the details of the extraction technique employed, we show how the leptodermous expansion of the smoothed HFB energy works.


\subsection{Pairing and Shell Corrections}
\label{Subsec:shell}

To extract  shell  and pairing 
corrections from the total HFB energy is not an easy task as the building blocks of HFB are quasi-particles, rather than the single-particle states that enter  
the Strutinsky energy theorem \cite{[Rin80]}. 
Moreover, while the contribution of pairing correlations to the 
total energy must be eliminated, the induced shape polarization must be 
kept, as it is relevant for making  the direct comparison with experiment.
 
To extract the effect of HFB pairing, we  carried out HF 
calculations  at the equilibrium  deformations of HFB g.s. and SD configurations. This was achieved by constraining 
the expectation values of the HF multipole moments 
at those corresponding to  HFB solutions.  In practice, it sufficed to consider 
$\hat{Q}_{20}$, $\hat{Q}_{40}$, $\hat{Q}_{60}$, and $\hat{Q}_{80}$ moments, higher-order terms being negligible. The advantages of  this procedure are twofold. Firstly, it enables us to  remove all direct pairing effects. Secondly, it provides a 
set of single-particle HF  states that can be used to compute the shell correction $\delta E_{\text{shell}}$.

Shell corrections were calculated according to the procedure outlined in Refs.~\cite{[Ver00],[Cwi05a]}. 
It combines the standard shell-correction  smoothing method (our original implementation is based 
on Ref.~\cite{[Bol72]}) with the Green's function oscillator expansion method
technique that is aimed at removing   the spurious contribution to $\delta E_{\text{shell}}$ stemming from the non-resonant continuum of positive  energy states. Following Ref.~\cite{[Kru00b]}, 
we employed the following smoothing parameters: smoothing widths  $\gamma_{n} = 1.66$ for neutrons and 
$\gamma_{p} = 1.54$ for protons (in units of $\hbar\omega_{0} = 41/A^{1/3}$) and 
the curvature correction $p = 10$. This choice guarantees that the generalized plateau condition is  satisfied \cite{[Ver00]}. 

\begin{figure}[h]
\includegraphics[width=\linewidth]{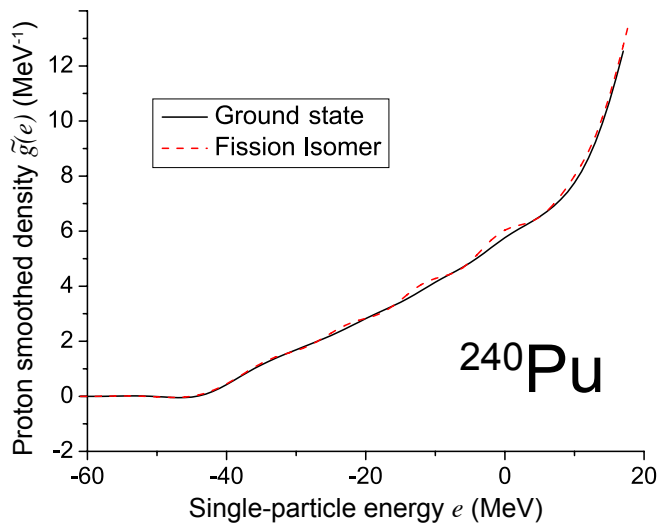}
\caption{
	     (color-online) Proton smoothed density $\tilde{g}(e)$ in the g.s. and SD configuration of $^{240}$Pu calculated with the SkM* EDF and $N_\text{max}=16$ HO shells.
	    }	 
\label{fig07}	 
\end{figure}

\begin{figure}[h]
\includegraphics[width=\linewidth]{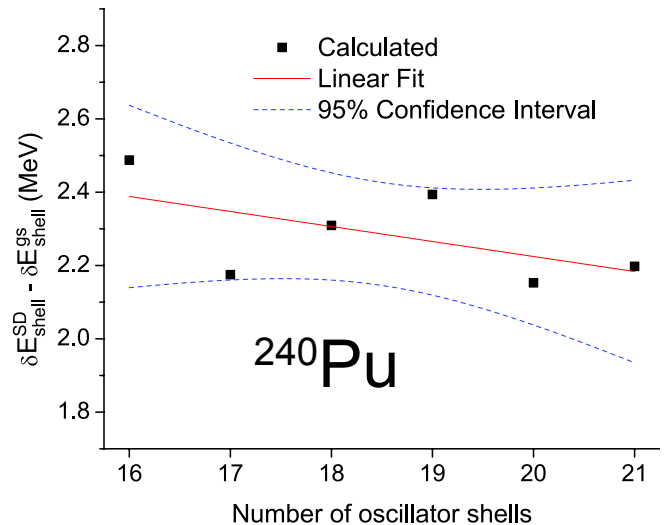}
\caption{
	     (color-online)  Convergence of the  shell correction 
	     contribution to the deformation energy, 
	     $\delta E_{\text{shell}}^{\text{SD}} - \delta E_{\text{shell}}^{\text{gs}}$, for  $^{240}$Pu
	     as a function of $N_\text{max}$. Calculations were performed with with the SkM* EDF. 
	    }	 
\label{fig07a}	 
\end{figure}

As an illustration,  Fig.~\ref{fig07} shows  the smooth 
density $\tilde{g}(e)$ for protons in the g.s. and fission isomer of $^{240}$Pu 
obtained with SkM* EDF.  The generalized 
plateau condition requires that this function must be linear across several oscillator shells, and this is indeed well fulfilled.
Figure~\ref{fig07a} displays the convergence of 
the   shell correction 
	     contribution to the deformation energy,
$\delta E_{\text{shell}}^{\text{SD}} - \delta E_{\text{shell}}^{\text{gs}}$, 
as a function of $N_\text{max}$. While the convergence is 
not perfect, the uncertainty remains fairly small, around 200 keV.

We should emphasize that there exist alternative  ways to extract  shell 
correction, see e.g., Refs.~\cite{[Sal10],[Pom04]}. However, since we use LDM parameters  extracted in Ref.~\cite{[Rei06]} by employing  the Green's function  prescription   \cite{[Ver00]}, it is important to follow the same procedure in order for our analysis to remain consistent.


\subsection{Determination of Microscopic LDM Deformations}
\label{Subsec:defs}

To compare the LDM deformation energy with the HFB bulk energy, 
one needs to  properly define the shape of  
the (sharp) surface  of the drop.
For the axial, reflection symmetric  shapes considered in this work, the drop surface is typically 
parametrized in terms of   deformations $\tilde{\beta}_{l}$ defining the multipole 
expansion of the drop radius. The problem consists therefore in mapping 
a set $\{ \tilde{\beta}_{l} \}_{l=2, 4,\cdots, N_Q}$ 
to a set of  multipole moments
$\langle\hat{Q}_{\lambda}\rangle^{\text{HF}}  =\langle r^{\lambda}Y_{\lambda 0}\rangle$ coming from  HF calculations. 

The LDM deformation parameters $\tilde{\beta}_{l}$ can be  determined from the  system of non-linear equations \cite{[Dud80b]}
\begin{equation}
\langle\hat{Q}_{\lambda}(\tilde{\beta}_{l})\rangle = \langle\hat{Q}_{\lambda}\rangle^\text{HF} \ \ \ \lambda = 0, 2, 4,\dots, N_{Q}.
\label{eq:Qexact}
\end{equation}
However, in such an approach involving standard (volume) multipole moments,  the role of  higher-order multipoles becomes artificially exaggerated. It was therefore argued (see Refs.~\cite{[Ner79],[Ben89a]} and references quoted therein)
that a mapping between the two sets of shape deformations can be best achieved by
using the surface multipole moments 
defined as $\hat{\mathcal{Q}}_{\lambda} \equiv r^{2}Y_{\lambda 0}$, which have a much softer radial dependence 
than volume  moments. Deformation parameters
$\tilde{\beta}_{l}$ can
therefore be extracted by requiring that 
the set of equations for dimensionless surface moments \cite{[Ner79],[Ben89a]}
\begin{equation}\label{surfdef}
\frac{\langle\hat{\mathcal{Q}}_{\lambda}(\tilde{\beta}_{l})\rangle}{\langle r^2(\tilde{\beta}_{l})\rangle} = \frac{\langle \hat{\mathcal{Q}}_{\lambda}\rangle^\text{HF}}
{\langle r^2 \rangle^\text{HF}}
\end{equation}
be satisfied, with $l, \lambda = 2, \dots, N_Q$. 
In practice only the  three lowest terms with $l= 2, 4, 6, 8$ are important at SD shapes. This choice provides  the best 
mapping between self-consistent  multipole moments and deformations of the sharp surface.


\subsection{Coulomb Polarization}
\label{Subsec:Cou}

In Ref.~\cite{[Rei06]}, the microscopic LDM parameters  were extracted from a set of spherical HF calculations without the Coulomb term. However, such a methodology is clearly not applicable in realistic calculations. First, the Coulomb term crucially affects nuclear deformability. Second, while its contribution to the total energy can easily be subtracted, the Coulomb potential induces a long-range  polarization of the mean-field, which affects the equilibrium deformations, single-particle states, etc. Most importantly, this Coulomb polarization is deformation-dependent. As a result, the contribution of the Coulomb term to the excitation energy, $E_{\text{Cou}}^{*}$, can vary by be up to several MeV for the interactions that we consider in this study.

To take this effect into account at the LDM level, we first extract the spherical charge radius $R_{0}^{\text{ch}} = r_{0}^{\text{ch}}A^{1/3}$ from the  self-consistent spherical HF calculations
for each   nucleus $(Z,N)$ 
and then use this value of $r_{0}^{\text{ch}}$ in Eq. (\ref{Eq_LDM_def}). 
Since the charge radius thus 
obtained does not come from a systematic fit but is obtained locally, it 
introduces shell fluctuations into LDM. However, 
since spherical self-consistent  radii behave smoothly as a function of shell filling  \cite{[Miz99]}
the corresponding shell effect is very small indeed. 

Our actual determination of the LDM charge radius goes as follows.  From the spherical  rms proton HF radius $\langle R_{p}^{2} \rangle$, we extract the rms charge radius $\langle R_{ch}^{2} \rangle$ according to the standard formula:
\begin{equation}
\langle R_{\text{ch}}^{2} \rangle = 
\langle R_{p}^{2} \rangle + 
\langle r_{p}^{2} \rangle +
\frac{N}{Z}\langle r_{n}^{2} \rangle,
\end{equation}
where $\langle r_{p}^{2} \rangle= 0.764$\,fm$^2$ and $\langle r_{n}^{2} \rangle=-0.116$\,fm$^2$ are the squared
rms  proton and neutron charge radius, respectively. The geometrical charge 
radius $R_{0}^{\text{ch}}$ is then  obtained from the rms charge radius 
in the usual way:
\begin{equation}
R_{0}^{\text{ch}} = \frac{5}{3}\sqrt{\langle R_{\text{ch}}^{2} \rangle}.
\end{equation}
We note that
the condition $(R_{\text{ch}}^2)^\text{LD} = \langle R_{\text{ch}}^{2} \rangle^\text{HF}$  together with Eq.~(\ref{surfdef}) implies that the charge surface of the microscopic liquid drop is close to that of HF.

\subsection{Example: Extraction of the Smooth Deformation energy for $^{236}$U}
\label{Uexample}

\begin{figure}[!]
\includegraphics[width=\linewidth]{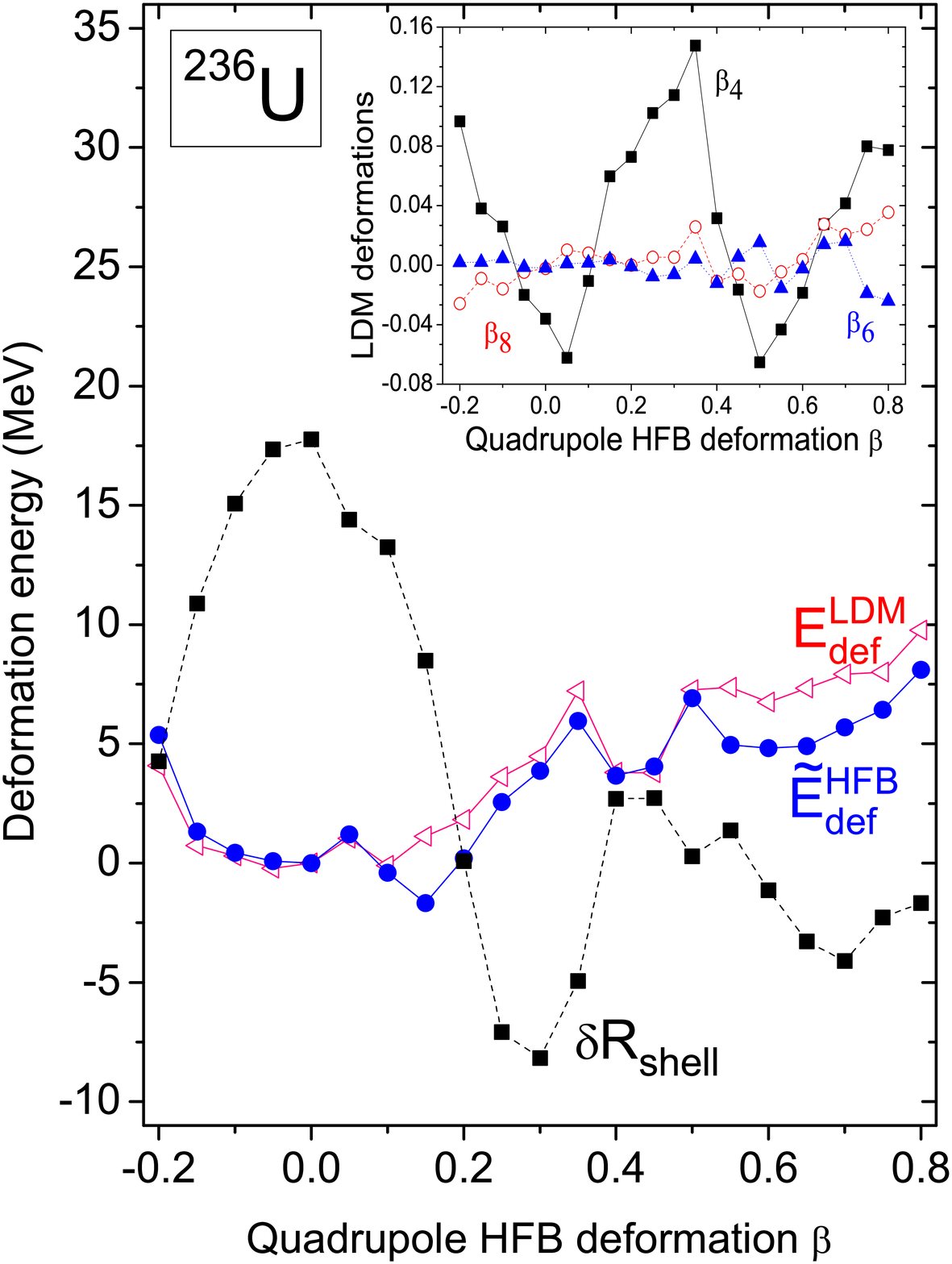}
\caption{
	(color-online) Extraction of the  LDM deformation energy from constrained HF+SLy4 calculations for $^{236}$U at several values  of quadrupole deformation $\beta$. Shown are: the total shell correction 
	$\delta E_{\text{shell}}$ (squares), 
	smooth HFB deformation energy 
	$\tilde{E}_{\text{def}}^{\text{HF}}(\beta)$ (dots),   and the corresponding LDM
	deformation energy $E_{\text{def}}^{\text{LDM}}(\beta)$ (triangles).
	The inset shows the equivalent LDM deformations $\tilde{\beta}_{l}$ with $l$=4,6, and 8.
	 }	 
\label{fig08}	 
\end{figure}

To illustrate the extraction procedure of smooth deformation energy from HF results, and to assess the quality of the leptodermous expansion, we  carry out constrained HF+SLy4 calculations for $^{236}$U shown in Fig.~\ref{fig08}.
The  constraint on the quadrupole moment was determined so that the deformation $\beta$ of Eq.(\ref{betadef}) takes the values 
$\beta = -0.20,-0.15,\dots,+0.80$. Since the HF potential energy curve consists of several sharply-crossing configurations as evidenced by rapidly varying LDM deformations shown in the inset of Fig.~\ref{fig08}, we made no attempt to interpolate between the mesh points in $\beta$. 
The shell correction was extracted at each $\beta$ according to the procedure 
discussed in Sec.~\ref{Subsec:shell}. The smooth energy at deformation $\beta$ is  given by:
$\tilde{E}^{\text{HF}}(\beta) = E^{\text{HF}}(\beta) - \delta E_{\text{shell}}(\beta)$, which defines
the smooth component of the HF deformation energy 
$\tilde{E}_{\text{def}}^{\text{HF}}(\beta) = \tilde{E}^{\text{HF}}(\beta) - \tilde{E}^{\text{HF}}(\beta = 0)$. At each $\beta$, the LDM deformation parameters $\tilde{\beta}_{l}$ are computed according to Sec.~\ref{Subsec:defs}.
Finally, the deformed LDM energy $E_{\text{def}}^{\text{LDM}}(\beta)$ is obtained using Eq.~(\ref{Eq_LDM_def})  with LDM constants taken from Table~\ref{table01} and the charge radius defined according to Sec.~\ref{Subsec:Cou}. As expected, the smooth deformation energy is growing
as a function of deformation; the local variations are due to configuration changes in HF calculations and corresponding changes in higher-order shape deformations.

Even though the agreement between $E_{\text{def}}^{\text{LDM}}$ and
$\tilde{E}_{\text{def}}^{\text{HF}}$ is not perfect, it is gratifying to see that the LDM energy nicely follows the smooth energy extracted from HFB. The deviation  has multiple sources 
such as the error on the regression analysis carried in \cite{[Rei06]},  uncertainties of the shell-correction procedure, neglect of the second-order effects in density fluctuations \cite{[Rin80]}, LDM assumption of the sharp surface, limitations of the leptodermous expansion used, etc. Considering all this, the quality of the leptodermous expansion of the deformation energy in deformed  nuclei is very reasonable.


\section{Surface-symmetry energy and deformed neutron-rich nuclei in DFT}\label{results}

We are now ready to determine the smooth part of the deformation energy from
HFB results and compare it with the microscopic LDM using the methodology described in Sec. \ref{Sec:leptodermous}.  In Fig.~\ref{fig03}  we demonstrated  that  the uncertainties in  the 
determination of $a_{\text{ssym}}$  impact the deformation 
properties of  neutron-rich nuclear drops.  It is therefore interesting to see  whether in the studies of well  deformed 
states in realistic nuclei  these uncertainties would show up.

\begin{figure}[htb]
\includegraphics[width=\linewidth]{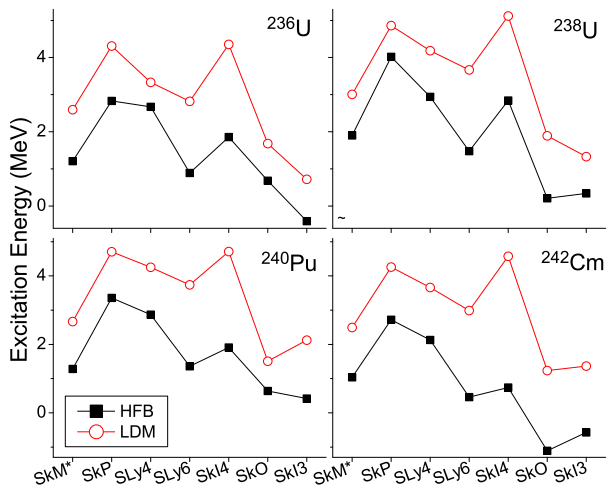}
\caption{
 (color-online) Smooth excitation energy $\tilde{E}^{*}$ of fission isomers in $^{236,238}$U, $^{240}$Pu, and $^{242}$Cu calculated in HFB and LDM for seven  EDFs. See text for details.
	    }	 
\label{fig09}	 
\end{figure}

Figure \ref{fig09} shows the smooth excitation energy $\tilde{E}^{*}$ of fission isomers calculated in HFB and LDM for seven  EDFs. For each nucleus, we first carried out HFB calculations to determine the g.s. and SD configurations. The constrained HF calculations are then performed  based on the multipole moments of the HFB solution. Shell energies are subtracted from g.s. and SD HF energies, and this defines the smooth part of the excitation energy $\tilde{E}^{*}_{\text{HFB}}$ in HFB, i.e., the smooth deformation energy of the excited state relative to the ground-state. Using the surface moments obtained in the g.s. and SD minima of HFB, we  extract the equivalent LDM deformation parameters $\tilde{\beta}_{l}$ and the LDM excitation energy.

\begin{figure}[htb]
\includegraphics[width=\linewidth]{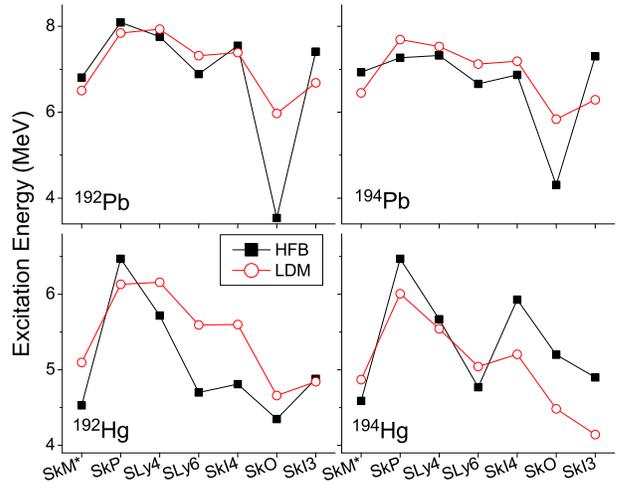}
\caption{
	     (color-online) Same as in Fig.~\ref{fig09} except for SD bandheads in 
	     $^{192,194}$Hg and  $^{192,194}$Pb.
	    }	 
\label{fig09a}	 
\end{figure}

As in Sec. \ref{Sec:HFB}, clear differences between various EDF parameterizations can be seen. Overall, these variations can be as large as 4 MeV at the LDM level. As discussed earlier in Sec.~\ref{Uexample}, there is a $\sim$2\,MeV shift of the LDM curves with respect to the HFB results. However, it is rewarding to see that the shift is systematic and the EDF-variations seen in HFB are properly captured by the equivalent LDM. The results for the SD bandheads are displayed in Fig.~\ref{fig09a}. In these lighter nuclei, the agreement between HFB and equivalent LDM is better on the average, but local fluctuations can be appreciable (see SkO or SkI3 results for Pb isotopes) and might be related to a complex pattern of g.s. equilibrium deformations in these nuclei.

The results shown in Figs. \ref{fig09} and \ref{fig09a}, combined with the overall picture of the residuals  in Fig. \ref{fig05}, demonstrate that large differences between Skyrme EDFs exist when it comes to deformation properties of nuclei. While these differences certainly depend on 
variations of EDF parameters controlling the shell structure, such as, e.g. the effective mass or spin-orbit splitting, our analysis indicates that there are also fundamental discrepancies at the level of the bulk energy. One may therefore question whether EDF optimization protocols 
based exclusively on a small amount of data in nuclear matter and spherical nuclei are able to capture the deformability of EDF. 

\begin{figure}[ht]
\includegraphics[width=\linewidth]{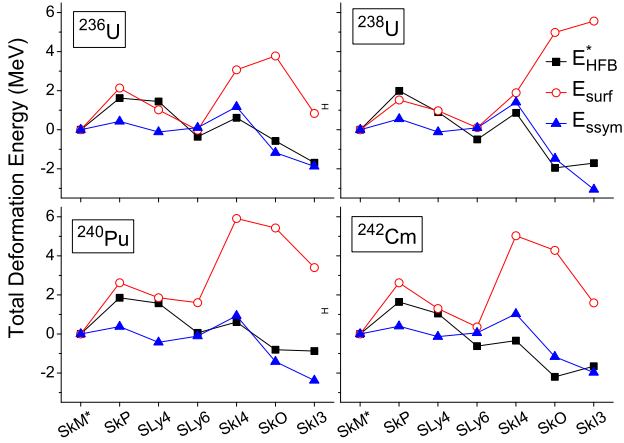}
\caption{
	     (color-online) Surface and surface-symmetry contributions to the LDM excitation energy of fission isomers in the actinides compared to  the smooth HFB excitation energy for  the same Skyrme EDFs
as in Figs. \ref{fig09} and \ref{fig09a}. All curves are shown relative to SkM* results.
	    }	 
\label{fig10}	 
\end{figure}

As discussed in Sec.~\ref{Sec:assym} and in particular in Figs.~\ref{fig01} and \ref{fig02}, $a_{\text{ssym}}$ varies very significantly from one EDF to another. Consequently, surface and symmetry properties of EDFs are intertwined in a non-trivial way when it comes to deformability. 
Guided by the results of Figs. \ref{fig09} and \ref{fig09a}, we may wonder whether the large variations in $a_{\text{ssym}}$  are indeed reflected in the results of self-consistent calculations.

\begin{figure}[!]
\includegraphics[width=\linewidth]{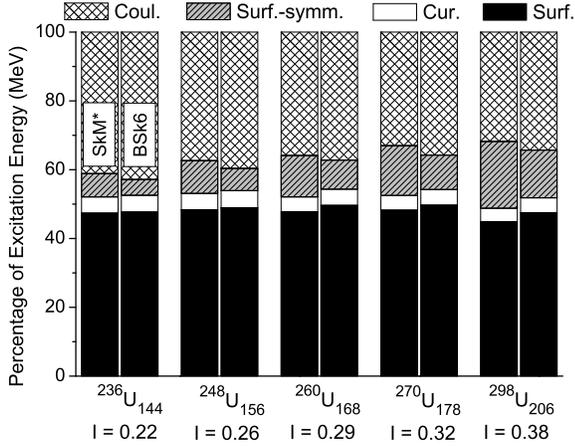}
\caption{
	     Relative contributions of the Coulomb, surface-symmetry, curvature and surface terms to the equivalent LDM excitation energy of SD states in a sequence of U isotopes. Calculations are based on SkM* and BSk6.
	    }	 
\label{fig11}	 
\end{figure}

Figure~\ref{fig10} shows the surface and surface symmetry contributions to the LDM excitation energy of SD states in the actinides for the same Skyrme EDFs
as in Figs. \ref{fig09} and \ref{fig09a}. The equilibrium deformations that are used in the LDM for both the ground-state and SD state are obtained in HFB. The LDM results are compared to the smooth HFB energy $\tilde{E}^{*}_{\text{HFB}}$. To facilitate interpretation, all curves are normalized to SkM* values. In this way, we can better compare relative variations obtained in various EDFs. It is interesting to see that the inter-EDF fluctuations of $\tilde{E}^{*}_{\text{HFB}}$ are rather well correlated with the surface-symmetry energy. In other words, the contribution from the Coulomb and curvature terms (not plotted in Fig. \ref{fig10} for better legibility) cancel out the surface term to a large extent. This result is significant because it seems to confirm the simple analysis of Sec. \ref{Sec:assym} in a realistic case: in nuclei having large neutron excess $I$ (here of the order of $I\approx 0.2$), differences in deformation energy between various EDF parameterizations reflect the differences of the surface-symmetry coefficient. By contrast, a similar analysis of individual macroscopic contributions in the Hg-Pb region does not allow to pin down a single LDM term as a primary deformation driver.

\begin{figure}[!]
\includegraphics[width=0.8\linewidth]{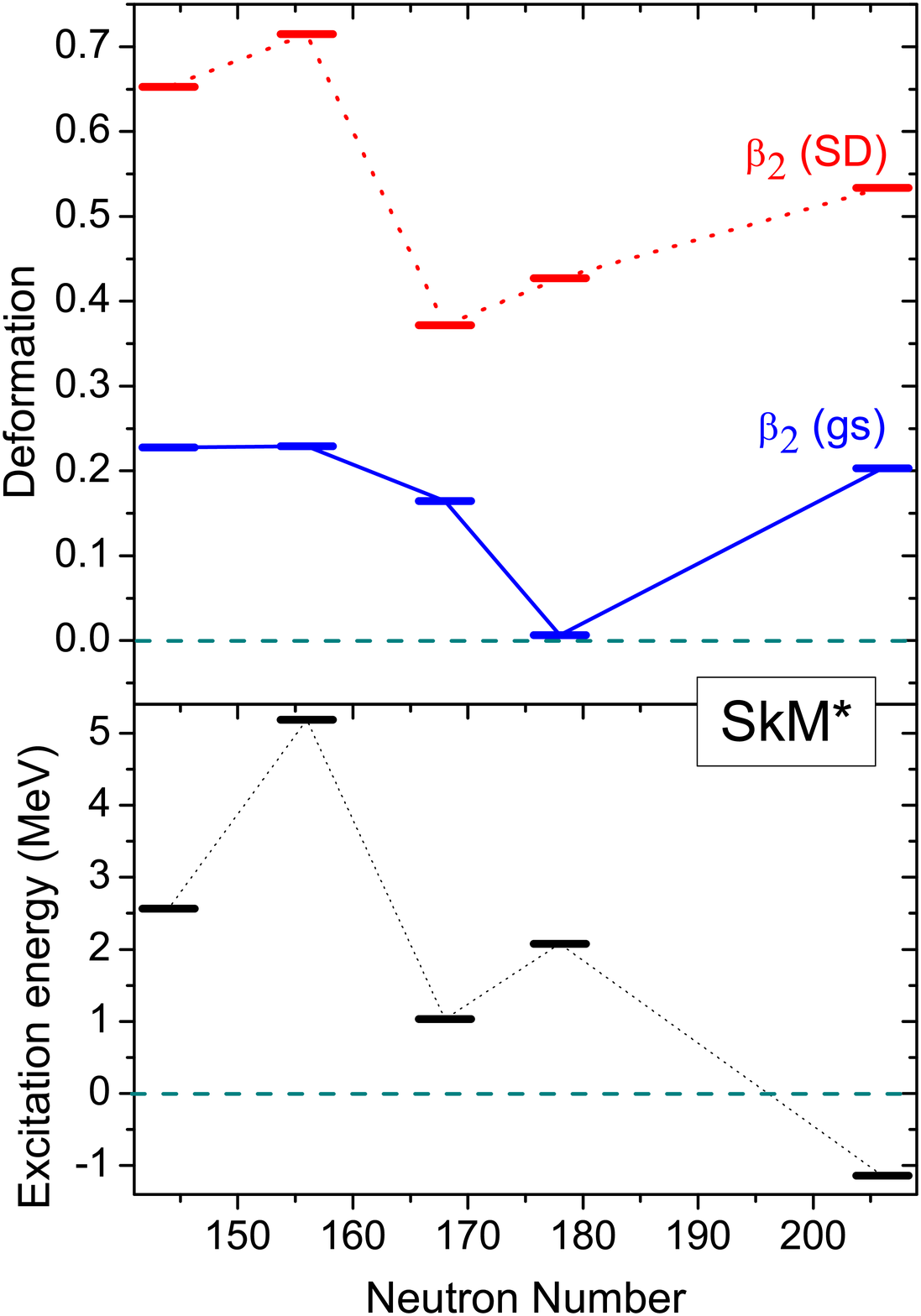}
\caption{
         Quadrupole deformations $\beta$  in SkM* for ground states and fission isomers for the same U isotopes as in Fig. \ref{fig11} (top)  and the corresponding excitation energies (bottom). 
	    }	 
\label{fig11b}	 
\end{figure}

To further illustrate the importance of the surface-symmetry term, we calculated the LDM excitation energy of the fission isomer for a sequence of U isotopes.
Here,  employed the SkM* and BSk6  parametrizations. The SkM* EDF is known to perform rather well for fission barriers \cite{[Bon04],[Sta09]}. Its surface-symmetry coefficient is also close to the average among Skyrme forces and values from phenomenological estimates, so it can be viewed as fairly representative of the Skyrme functionals. The BSk6 parameterization gives a reasonable rms deviation for excitation energies of SD states, see Fig.~\ref{fig05}.  The isotopes considered include some very neutron-rich species important in the context of r-process fission recycling. It is worth noting that HFB potential energy landscapes change considerably within this isotopic sequence. For example, $N$=184 is a neutron magic number for SkM* \cite{[Kru00b]}; hence, g.s. configurations around $^{276}$U are spherical rather than prolate, see Fig. \ref{fig11b}. The equivalent LDM equilibrium deformations, therefore, reflect these structural changes in a non-trivial way.

For each isotope, we computed the relative contribution of the surface, surface-symmetry, curvature, and Coulomb term to the total LDM excitation energy: these are the only terms that depend on deformation. Figure \ref{fig11} shows the percentage of the LDM excitation energy from these contributions. As the neutron excess grows, one can notice the gradual relative decrease of the Coulomb contribution - which depends only on the number of protons and proton density, and therefore remains relatively constant in value. This decrease is compensated by an increase of the surface-symmetry contribution. For the most neutron-rich nuclei considered here, the total contribution from $E_{\text{ssym}}$ is as large as 19\% for  SkM*. For the two parametrizations selected here, the role of the surface-symmetry term increases by a factor 3 from $^{236}$U to the fission recycling region. 


\section{Conclusions}
\label{conclusion}

This work contains a comprehensive study of deformation properties of nuclear energy density functionals based on the leptodermous expansion of the smooth nuclear energy. Since symmetry and surface-symmetry 
terms in the expansion are strongly correlated, a way to resolve them is to consider data on deformed neutron-rich nuclei, in which the surface-symmetry term is amplified. Based on intuitive LDM arguments, we argue that deformation 
properties of neutron-rich nuclear drops are governed by an interplay of the 
deformation-driving Coulomb and surface-symmetry terms, and the surface energy that acts against shape deformation. To estimate this interplay, we extracted the smooth deformation part of the HFB energy by means of the shell correction procedure. 

Self-consistent DFT calculations for excitation energies of SD states show  marked differences in  their predictions   depending on the parametrization used. 
For the set of EDFs considered,  the average rms deviation between predicted 
energies of SD states and experimental values is 1.26\,MeV. Within this set, the MSk1 parametrization provides the best overall reproduction of the data: the corresponding  rms deviation is  0.53\,MeV, and this set a benchmark for future improvements. For the subset of fission isomer data, the best performer is SkI4: its rms deviation is 0.48\,MeV.

We showed that inter-parametrization differences reflect to a large extent  macroscopic 
properties of EDFs. In particular, our calculations indicate
 that the bulk deformation properties of actinides are strongly  driven 
by surface-symmetry effects, while in the proton-rich $A\sim 190$ nuclei there is more competition between the various macroscopic contributions. 
This finding  should have an impact on the fissility of heavy, very neutron-rich 
nuclei of the kind encountered e.g. in the r-process. For example, 
the surface-symmetry contribution to the bulk part of the excitation energy of 
fission isomer in very neutron-rich uranium isotopes  can reach  $\sim$20\% as  compared with $\sim$5\% for $^{236}$U. 

\begin{figure}[h]
\includegraphics[width=0.5\textwidth]{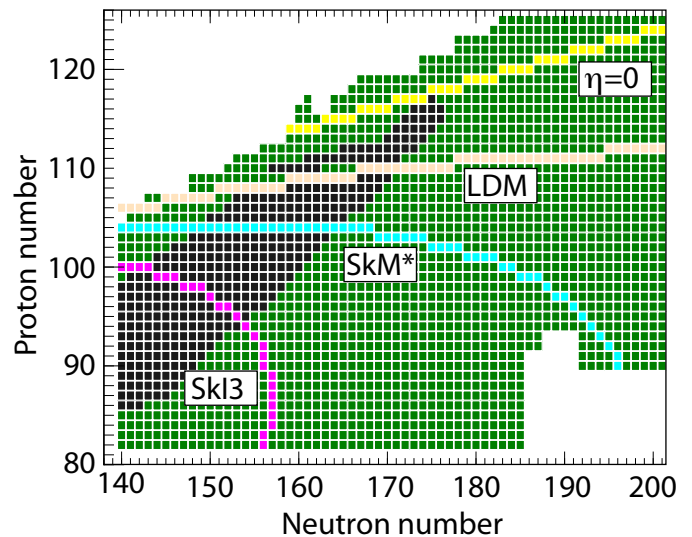}
\caption{The upper part of the chart of the nuclides with the $x=1$ limit indicated for SkI3 and  SkM* EDFs, the value of  $\eta=1.7826$  used in Ref.~\cite{[Bel67]} (LDM), and no isospin dependence ($\eta=0$). The region of known nuclides is marked by black squares.
	    }	 
\label{figfiss}	 
\end{figure}

The importance of the surface-symmetry term on fission can be quantified at the LDM level by the dimensionless fissility parameter:
\begin{equation}\label{fissility}
x = \frac{E_\text{Coul}(\text{sph})}{2E_\text{surf}(\text{sph})}\approx \frac{Z^2}{47 A\left(1-
\eta I^2
\right)},
\end{equation}
where $\eta\equiv -a_{\text{ssym}}/a_{\text{surf}}$.
If $x>1$, the nuclear liquid drop is unstable to fission. In the presence of neutron excess, the fissility parameter increases, i.e., the tendency towards fission increases. 
In Refs.~\cite{[Bel67],[Cam02]}, the value $\eta=1.7826$ was used. By taking 
LDM parameters from Table~\ref{table01} we see that $\eta$ is 1.9 for BSk6, 2.9 for SkM*, and 4.16 for SkI3, i.e., this parameter is very uncertain. 

Figure~\ref{figfiss} shows the LDM fission limit for the SkI3 and  SkM* EDFs, as well as for    
$\eta=1.7826$, and $\eta=0$ (no isospin dependence). The minimum value 
obtained for $Z^2/A$, i.e., 47 in Eq.~(\ref{fissility}), is not very precise as it depends on assumptions about the LDM constants \cite{[Den03]}. Therefore
this diagram should be considered as a qualitative guidance. A clear message drawn from Fig.~\ref{figfiss} is that the surface-symmetry term can significantly impact LDM fission barriers: the greater the value of $\eta$, the lower the threshold for fission. This result is especially important in the context of the fission recycling mechanism in the r-process and hot fission reactions leading to excited neutron-rich  superheavy nuclei. Since shell effects are to a large extent washed out at high temperatures \cite{[Bra72],[Mor72]}, the fission of hot compound nuclei is expected to be governed by the LDM fission barrier (or smooth HFB deformation energy). As seen in Fig.~\ref{figfiss} the uncertainty in $a_{\text{ssym}}$, hence $\eta$, makes it difficult to reliably predict  fission rates of the heaviest and superheavy neutron-rich nuclei. (In this context we note that according to the recent estimates \cite{[Lee10]} $a_{\text{ssym}}$ and $\eta$ are expected to very weakly depend on temperature.)

The results obtained in this paper suggest that adding to the list of fit-observables data on strongly deformed nuclear states
 (such as excitation energies of SD states or fission barriers), 
combined with the usual  constraints on bulk properties and   shell structure, should  constrain quite effectively the surface properties of the nuclear EDF. Such a  strategy is currently being pursued  within the UNEDF project \cite{[UNEDF],[Ber07a]}. On the experimental side, new information on deformed properties on neutron-rich systems is the key.

\begin{acknowledgments} 
Useful discussions with J. Skalski, A. Staszczak,  M. Stoitsov, and P.G. Reinhard are gratefully appreciated. This work was supported by the U.S. Department of Energy under Contract Nos. DE-FC02-09ER41583 (UNEDF SciDAC Collaboration), DE-FG02-96ER40963 (University of Tennessee); by the National Nuclear Security Administration under the Stewardship Science Academic Alliances program through DOE Grant DE-FG52-09NA29461; and by the NEUP grant DE-AC07-05ID14517 (sub award 00091100). This work was partly performed under the auspices of the US Department of Energy by the Lawrence Livermore National Laboratory under Contract DE-AC52-07NA27344. Computational resources were provided by the National Center for Computational Sciences at Oak Ridge National Laboratory.
\end{acknowledgments}


\bibliographystyle{unsrt}

\end{document}